\documentstyle[12pt,twoside,fleqn,espcrc1,epsfig]{article}

\newcommand{\AmS}{{\protect\the\textfont2
  A\kern-.1667em\lower.5ex\hbox{M}\kern-.125emS}}

\title{Hadronic Expansion Dynamics in Central Pb+Pb Collisions 
at 158 GeV per Nucleon} 

\def\be{\begin{equation}}
\def\ee{\end{equation}}
\def\bea{\begin{eqnarray}}
\def\eea{\end{eqnarray}}

\newcommand{\qinv}{\mbox{$Q_{inv}$}}

\newcommand{\ycm}{\mbox{$y_{cm}$}}
\newcommand{\ypi}{\mbox{$y_\pi$}}
\newcommand{\ypipi}{\mbox{$Y_{\pi\pi}$\ }}
\newcommand{\kperp}{\mbox{$K_{\bot}$\ }}
\newcommand{\Rzero}{\mbox{$R_0$\ }}
\newcommand{\Rperp}{\mbox{$R_{\bot}$\ }}
\newcommand{\Rpar}{\mbox{$R_{\parallel}$\ }}
\newcommand{\yYKP}{\mbox{$Y_{YKP}$\ }}
\newcommand{\Betayk}{\mbox{$\beta_{YK}\ $}}

\begin{document}
\begin{center}
{\large {\bf Hadronic Expansion Dynamics in Central Pb+Pb Collisions 
at 158 GeV per Nucleon}} 

\sloppy
\vspace{0.5cm}
\noindent
{\large NA49 Collaboration}
\end{center}
H.~Appelsh\"{a}user$^{7,\#}$, 
J.~B\"{a}chler$^{5}$, S.J.~Bailey$^{16}$, L.S.~Barnby$^{3}$, J.~Bartke$^{6}$, 
R.A.~Barton$^{3}$, H.~Bia{\l}kowska$^{14}$, A.~Billmeier$^{10}$, 
C.O.~Blyth$^{3}$, R.~Bock$^{7}$
% ...

\vspace{0.5cm}
\noindent
% ...
$^{3}$Birmingham University, Birmingham, England,
% ...
$^{5}$CERN, Geneva, Switzerland,
$^{6}$Institute of Nuclear Physics, Cracow, Poland,
$^{7}$Gesellschaft f\"{u}r Schwerionenforschung (GSI), Darmstadt, Germany,
% ...
$^{10}$Fachbereich Physik der Universit\"{a}t, Frankfurt, Germany,
% ...
$^{14}$Institute for Nuclear Studies, Warsaw, Poland,
% ...
$^{16}$Nuclear Physics Laboratory, University of Washington, Seattle, WA, USA,
% ...
$^{\#}$present address: Physikalisches Institut, Universitaet Heidelberg, Germany.

% ...

\newpage 

\begin{abstract}
{\bf Abstract}\\

Two-particle correlation functions of negative hadrons over wide 
phase space, and transverse mass spectra of negative hadrons and deuterons near
 mid-rapidity have been measured in central Pb+Pb collisions at 158
GeV per nucleon by the NA49 experiment at the CERN SPS. 
A novel Coulomb correction procedure for the negative two-particle 
correlations is employed making use of the measured oppositely charged 
particle correlation.
Within an expanding source scenario these results are used to extract 
the dynamic characteristics of the hadronic source, resolving the ambiguities 
between the temperature and transverse expansion velocity of the source,
that are unavoidable when single and two particle spectra are analysed
separately.
The source shape, the total duration of the source expansion, the duration 
of particle emission,
the freeze-out temperature and the longitudinal and transverse expansion 
velocities are deduced.

\end{abstract}

\section{Introduction}

Lattice QCD calculations \cite{Laer96} predict that a phase transition
occurs between hadronic and deconfined, chirally symmetric partonic 
matter at energy densities of
about 2~GeV/fm$^3$, corresponding to a critical temperature $T_{c}$ 
between 150 and 200 MeV in baryon-free matter. 
Studies of transverse
energy distributions in central Pb+Pb collisions at the CERN-SPS with
158~GeV per nucleon $^{208}$Pb beams ($\sqrt{s}\approx 17$~GeV per
nucleon pair) show that energy densities of 2-3~GeV/fm$^3$ are
created in the initial interaction volume \cite{Albe95}. It is therefore
conceivable that such a transition is induced in ultrarelativistic
nuclear collisions. The partonic phase created in the initial stage of
the collision hadronizes as the system expands and cools. 
The expansion dynamics of the final hadronic phase are expected to
differ considerably with the presence or absence of a prior plasma phase, 
and it is suggested that the observed final state hadrons 
reflect the space-time evolution of the system and therefore 
provide clues as to the state of the system prior to hadronization.

The space-time evolution of a pion emitting source can be
probed using the momentum-space correlation of identical pions. 
Bose-Einstein correlations have been first used to study the
properties of a hadron emitting system created in hadronic collisions
under the assumption of a static source \cite{Gold60}.
The expansion of a source that is not static but
exhibits a hydrodynamical scaling evolution has been discussed by
Shuryak and Bjorken
\cite{Shur80,Bjor83}.
Within the framework of such models
it has been further shown that the dependence of
the correlation function on the pion pair kinematic variables is
an important diagnostic tool for the study of an expanding system 
\cite{Siny89,Bert89,Hein96l}.

Experiments NA35 \cite{NA35a} and NA44 \cite{NA44a} have applied this technique
to the investigation of central collisions of $^{32}$S nuclei with heavy
nuclear targets at the CERN SPS, resulting in the picture of a
cylindrically symmetric source expanding predominantly in the
longitudinal direction, with some indication of concurrent
transverse expansion. NA35 has observed that the incoming energy is
not completely stopped for these systems \cite{Bach91}, consistent with the
predominance of longitudinal expansion. The longitudinal evolution was
found to be consistent with Makhlin and Sinyukov's analysis 
\cite{Siny89} of ``scaling'' expansion, 
which is based on Bjorken's earlier study \cite{Bjor83} of an
idealised hydrodynamic longitudinal expansion model.
Longitudinal collective expansion is not unique to relativistic 
nucleus-nucleus collisions. 
However the transverse expansion dynamics, seen through
 single particle and two particle observables, is expected to be sensitive to
 the 
existence of a first order hadronization transition \cite{Bert89}, 
as well as to
the occurrence of a near isentropic hadronic expansion \cite{Schn93}
which would preserve information about the initial phase, and the existence 
of a primordial partonic transverse expansion velocity \cite{Fere96,Herm95}.

Transverse collective expansion is expected to be more prominent in the
heavy Pb+Pb collision system than in collisions of lighter
projectiles,
owing to the smaller surface-to-volume ratio, the larger ``fireball'' volume
and the longer duration of expansion. 
Indication that this is the case is provided by the
single particle transverse mass spectra of kaons,
protons and $\Lambda$s produced in central Pb+Pb collisions at the
CERN SPS \cite{Bear96,Afanas}, which exhibit unusually high inverse 
slope parameters (``temperature''),
in excess of 250 MeV, ruling out a simple thermal hadronic gas
model \cite{Hage83}. These distributions can be reproduced by 
invoking a radial velocity field in transverse direction, 
though the single particle spectra alone cannot distinguish between an elevated
temperature and a boost due to transverse collective expansion 
\cite{Schn93,Dodd95}
and therefore cannot determine uniquely the temperature
and transverse velocity.

It will be shown that
the combination of single particle spectra with two-particle correlation
results can lift this essential ambiguity and determine both the expansion
velocity and temperature at freezeout.
The term freezeout refers here to the final decoupling of hadrons
from the strong interaction. Note that the term ``freezeout''  is often 
also employed for 
hadronic decoupling from inelastic interactions at which point the population
ratios of hadronic species become stationary \cite{Soll94}.
This stage obviously precedes the one under study here. 

The analysis of this paper is based upon measurements by the NA49 experiment of 
two-particle correlation functions of negative hadrons over a wide
phase-space region and on single particle transverse mass spectra of negative
hadrons and deuterons near and below mid-rapidity, for central Pb+Pb
collisions at 158 GeV per nucleon. 
The like-charge pion correlation functions are corrected for Coulomb effects
using a novel technique based upon the parametrization of unlike-charge pion 
correlation functions measured in the same experiment \cite{Albe97}. 
The corrected like-charge correlation functions are fitted with the Yano-Koonin-Podgoretskii
parametrization \cite{YKP78} as formulated by Heinz \cite{Hein96} 
and the extracted source parameters are studied as
function of rapidity and transverse momentum of the particle pair. 
In the framework of an expanding hadronic source model, 
such an analysis of correlation
functions yields information about the source size and the duration of
the freezeout process, as well as the ratio ${\beta_{\bot}^2}/T$ near
mid-rapidity, where $\beta_{\bot}$ is the transverse velocity
and $T$ is the freezeout temperature. The ambiguity between
temperature and expansion velocity implicit in the ratio can then be  
lifted by considering single particle transverse mass spectra. A consistent
picture of collective velocity fields governing 
longitudinal and transverse expansion emerges,
with an estimated freezeout temperature at mid-rapidity of 120 MeV 
and a transverse expansion velocity of  $\beta_{\bot}$ = 0.55.

The paper is organized as follows: section \ref{sec:experiment}
describes the experimental procedure and data analysis, section
\ref{sec:corr_funct} gives a discussion of the experimental
determination of correlation functions, section \ref{sec:sourceparams}
describes the determination of the source parameters through fitting
with the Yano-Koonin-Podgoretskii parametrization of the correlation
functions, section \ref{sec:sourceparamdisc} discusses the extracted
parameters in terms of source size, average freezeout time and duration,
freezeout temperature and expansion velocity, section \ref{sec:spectra}
introduces transverse mass spectra to determine uniquely the freezeout
temperature and expansion velocity within the framework of a
hydrodynamical model, and conclusions are drawn in 
section~\ref{sec:conclusions}.

\section{Experimental Procedure and Data Analysis} 
\label{sec:experiment} 

\begin{figure} 
\epsfig{figure=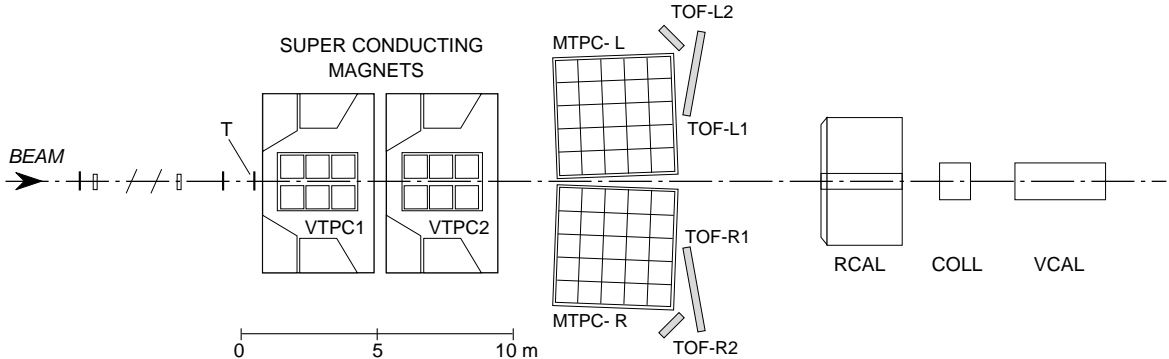,height=4.8cm}
\caption{
Schematic diagram of the NA49 experiment. The beam entering from the left 
interacts with the target T, producing charged particles most of which
are analyzed 
by two superconducting magnets and their trajectories are recorded by four 
Time Projection Chambers (TPCs) with an active volume of 50 m$^3$. Particle 
identification is done by time of flight and dE/dx. A set of calorimeters 
cover the forward rapidity region and are used for triggering purposes. 
\label{fig:na49}
}
\end{figure}

The NA49 experiment \cite{Afanas}, shown schematically in
Fig.~\ref{fig:na49}, is a large acceptance hadron spectrometer at
the CERN SPS. It comprises four large Time Projection Chambers (TPCs),
four Time-of-Flight walls (TOF),
and a forward (Veto) calorimeter for triggering. 
The target (T) was a foil of natural lead with thickness 224 mg/cm$^2$
(1\% interaction probability for the $^{208}$Pb beam ), mounted 80 cm 
upstream of the first TPC.

Two of the TPCs (VTPC1 and VTPC2) are placed inside dipole
 magnets
of 9 Tm total bending power 
and achieve a momentum resolution ${\delta}p/p^2$ = 10$^{-4}$ GeV$^{-1}$ .
The TPCs downstream of the magnets (MTPCs), extend the acceptance 
to high momentum and provide
particle identification via ionization measurement 
with a resolution of 4.5\% in $dE/dx$. 
The rapidity coverage of the TPCs is $2 < y_{\pi} < 5.5$, and the
transverse momentum range selected for this analysis
is $0.05 < p_T < 2$~GeV/$c$, with
almost complete azimuthal coverage. The Time-of-Flight walls
cover a rapidity of $2 < y_{\pi} < 3.5$ with an average time
resolution of about 60~ps and a flight path of about 14 m.  

The particle
identification using the measurement of the
specific ionization in the TPC gas in the relativistic rise regime
is complemented by the time-of-flight measurement at the lower momenta. 
In the present work we utilize particle
identification only for the deuteron transverse mass spectra. 
Two-particle correlations and pion transverse mass spectra are based upon
measurement of negative hadrons, which are dominantly pions (see
below).

A prompt trigger selects events from the lowest 3\% of the forward
energy distribution detected in the Veto calorimeter, located 27 m
downstream of the target and covering the beam fragmentation region
(laboratory rapidity $ y \geq 5.5$). This anticoincidence trigger,
which vetoes events with a large number of projectile spectator
nucleons, serves as a ``central collision'' trigger selecting 
impact parameters less than 3 fm \cite{Albe95}. About 1,200 charged
particles are recorded in the NA49 TPCs for each central Pb+Pb collision.

Negative hadron two-particle correlation functions are studied using
80,000 events analyzed independently in the second Vertex TPC and in
the Main TPCs. Negative hadron and deuteron transverse mass spectra
are analyzed using 300,000 events in which charged particles are tracked 
from the MTPCs to the TOF walls. 
The detailed techniques employed for cluster finding, tracking,
efficiency calibrations, non-vertex track rejection, and hadron
identification \cite{Appe97} will be described in a forthcoming
publication.

Since the two-particle correlation function carries the information of the 
source size in the small relative pair momentum range, 
and this region becomes narrower in momentum space 
the larger the source size is, the relevant 
signal for central Pb+Pb collisions is expected to occur at momentum
differences less than 50 MeV/$c$ in the pair centre of mass frame.
Therefore  the required resolution for
the measurement of the momentum difference of two tracks is less
than about 10 MeV/$c$. The momentum resolution was studied using a Monte
Carlo simulation chain based on the event generator
VENUS \cite{Wern88} and the detector response simulation code GEANT
\cite{Brun15}. 
The resolution in all relative momentum projections was shown to be
($5\pm1.5)$ MeV/$c$ throughout the acceptance used in this study,
$2.9<\ypipi<5.5$ and pair mean transverse momentum $\kperp<0.6$~GeV/$c$
\cite{Appe97}.

\section{Discussion of Correlation Functions}
\label{sec:corr_funct}

Experimentally, the normalized two-particle correlation function is
obtained from the ratio

\begin{equation}
\label{eq:exp_corr_fn}
C_2(Q) = \frac{A(Q)}{B(Q)} 
\end{equation}

\noindent where $A(Q)$ is the measured two particle probability
distribution as a function of relative four momentum $Q = p_1-p_2$, 
and $B(Q)$ is the
(uncorrelated) background distribution, calculated in the same way but
using two particles taken from different events. 

The dynamical expansion of the source in central collisions
is reflected in the dependence
of the correlation function on all five non-trivial momentum components 
of the pair kinematics, in which the azimuthal angle with respect to the 
beam axis is averaged out. For extraction of the source parameters from
the correlation functions we employ the Gaussian approximation in a representation given by Heinz
that is appropriate to explicitly reveal longitudinal collective 
motion \cite{Hein96l,Hein96}:

\begin{eqnarray}  
C_2(Q_{\bot}, Q_{\parallel}, Q_0; \ypipi, \kperp) & = & 1+\lambda
exp\:[-Q^2_{\bot}R^2_{\bot}-\gamma_{YK}^2(Q_{\parallel}-\beta_{YK}Q_0)^2R^2_{\parallel}-
\nonumber\\ 
& & \gamma_{YK}^2(Q_0-\beta_{YK}Q_{\parallel})^2R^2_0]. 
\label{eq:ykph_corr_fn}
\end{eqnarray}

\noindent 
This expression is an explicit function of the energy difference
$Q_0 = E_1-E_2$, transverse momentum difference 
$Q_{\bot} = ((\Delta p_x)^2+(\Delta p_y)^2)^{1/2}$, 
longitudinal momentum difference
\mbox{$Q_{\parallel} = \Delta p_z$}, and correlation intensity parameter 
$\lambda$, and an
implicit function of the pair rapidity

\begin{equation}
\label{eq:ypipi}
Y_{\pi\pi} = \frac{1}{2}\ln(\frac{E_1+E_2+p_{z1}+p_{z2}}{E_1+E_2-p_{z1}-p_{z2}}).
\end{equation}

\noindent
and the mean transverse momentum 
$\kperp = \frac{1}{2} \sqrt{(p_{x1}+p_{x2})^2 + (p_{y1}+p_{y2})^2}$. 
$\beta_{YK}$ is the ``Yano-Koonin velocity'' which
describes the source's longitudinal collective motion
in each interval of \ypipi. 
A non-zero $\beta_{YK}$ results in mixing the
longitudinal momentum and energy differences, $Q_{\parallel}$ and $Q_0$,
in determining  the longitudinal source parameter $R_{\parallel}$ and the time
duration parameter $R_0$. 
The parameters $\lambda,\:R_{\bot},\:R_{\parallel},\:R_0$ and $\beta_{YK}$ are
extracted from a fit to the correlation function data.
The parameters $\:R_{\bot},\:R_{\parallel}$ and $\:R_0$ 
describe the apparent source extent in transverse, longitudinal and temporal
directions. They still depend on pair rapidity \ypipi and pair transverse
momentum \kperp. 
This dependence will be further analysed in sections 4 and 5, resulting
in the extraction of the actual geometric space-time parameters
of the source within an explicit model with longitudinal and transverse expansion 
\cite{Hein96l}. 

Due to its large acceptance, NA49 was able to determine the dependence
of expression~(\ref{eq:ykph_corr_fn}) on all five kinematic quantities
over a wide region of \ypipi and \kperp, giving a complete picture of
the source expansion dynamics.

To investigate the effect
of the finite resolution and other experimental effects on the
correlation functions, a simulation of the Bose-Einstein
correlations \cite{Appe97}  was applied to the
(uncorrelated) VENUS events, which were then processed through the
detector simulation and the same analysis chain that was applied to the
data. In this way, the response of the device was estimated to generate
a lowering of the source parameters $R_0,\:R_{\bot}$ and
$R_{\parallel}$ by an amount between 2 and 8\%. No correction for this
effect was applied to the parameters reported below.

The correlation functions were calculated using all negative hadrons
rather than identified $\pi^-$ pairs. The Monte Carlo simulation chain
showed that the $h^-h^-$ pair signal consists of about 55\% genuine
$\pi^-\pi^-$ pairs from the primary event vertex, with this fraction
uniform over the measured phase-space. The remaining background pairs
arise principally from three sources: (i) $K^-$, electron and
antiproton contamination from the primary interaction; (ii) $\pi^-$ from
$K$ and $\Lambda$ decays in flight that were erroneously flagged as
primary vertex tracks, and (iii) secondary tracks generated in the
detector material that could not be removed by primary vertex
cuts. However, simulation of the correlation functions comparing the
$h^-$ pair signal to the genuine $\pi^-$ pair signal showed that the
contamination is ``well behaved'', in that it contributes uniformly
to all two-track relative momenta in all projections of the
correlation function. As a result, the background contamination 
principally affects the absolute magnitude of the correlation signal
but not the shape of the correlation functions. In an analysis
employing equation (\ref{eq:ykph_corr_fn}), the Monte Carlo simulation 
thus showed that only the $\lambda$ parameter, which quantifies the overall
correlation strength falls from its ideal value of unity down to
$\lambda\approx0.5$ due to contamination in the true $\pi^-$ pair
signal. The source parameters $R_0,\:R_{\bot}$ and
$R_{\parallel}$, which depend upon the shape of the correlation
function, deviate at the level of only 2 to 6\%. No correction for
this effect was applied to the parameters reported below.
Due to the large acceptance, high multiplicity and large number
of analysed events, the statistical errors on the correlation
functions are negligible.  Taking into account the systematical error
of the fit procedures, we estimate the systematic errors of space-time
parameters extracted from the correlation function to be $\pm 15\%$.
 
In order to extract the correlation function for
identical charged pions which is due to quantum statistical effects,
and which is sensitive to the characteristics of the emitting source,
it is necessary first to correct the observed correlation for the
effect of Coulomb repulsion. 
The distribution in relative momentum of two charged pions is strongly
affected by Coulomb repulsion as their momentum difference, $Q$, 
approaches zero. 
The correction for the Coulomb repulsion is done by applying a weight to
each pion pair in the background distribution $B(Q)$ in equation
(\ref{eq:exp_corr_fn}). This weight has traditionally been the inverse
of the Gamow penetrability factor \cite{NA35a}. The Gamow factor is
``pointlike'', in that it describes the emission of an isolated pair
of charged particles from a small and otherwise neutral spatial region
($r<1\:$fm). Because of the finite size of realistic sources \cite{Baym96}
and the multiparticle Coulomb screening which occurs when a large
number of charged pions are simultaneously emitted \cite{Albe97},
Gamow factor corrections are expected not to be appropriate for central 
Pb+Pb collisions at CERN SPS energies.

\begin{figure} 
\begin{center}
\epsfig{figure=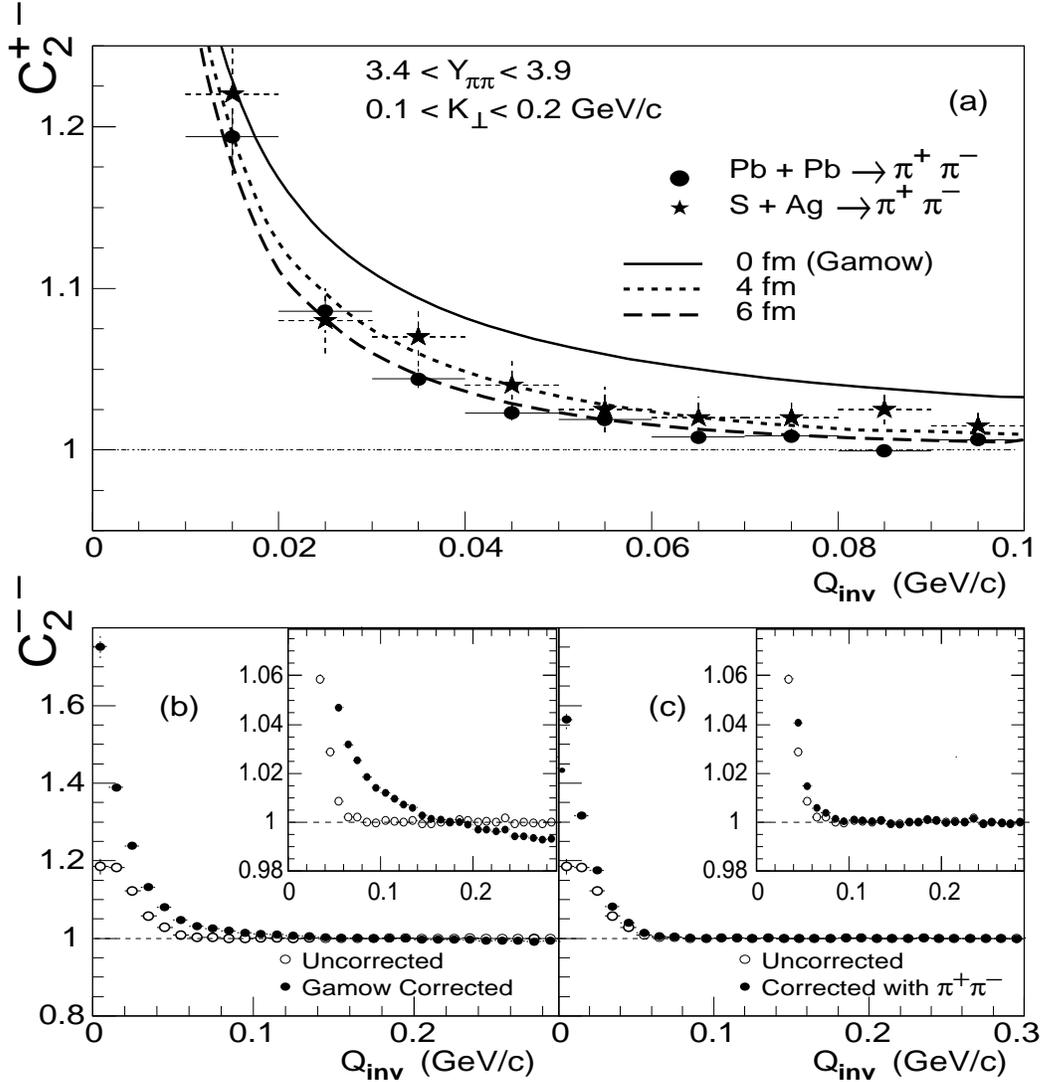,width=14.0cm}
\end{center}
\caption{
Panel (a): Correlation function of pairs of oppositely charged pions
as a function of \qinv, for central collisions of S+Ag (stars) and
Pb+Pb (circles), within the indicated acceptance. Also shown are
calculated correlation functions based upon the Coulomb interaction
for a point-like source (solid line) and finite-sized sources of
radii 4 fm (dotted line) and 6 fm (dashed line) \cite{Baym96}. Panel (b):
Correlation function of pairs of negatively charged pions as a
function of \qinv, for central collisions of Pb+Pb, within the same
acceptance as panel (a). Open circles are uncorrected data, filled
circles are data corrected for Coulomb repulsion calculated with
standard Gamow function for point-like source. The insert shows the same
distribution with expanded vertical scale. Panel (c): same as panel
(b), but correction now based upon finite-sized source of radius 6
fm \cite{Baym96}.
The correlation functions have been derived from positive and negative hadron
data by an appropriate Monte-Carlo correction for non-pionic contaminations.
\label{fig:PlusMinus}
}
\end{figure}

The Coulomb correction for charged pions can be investigated
experimentally through measurement of the correlation functions of
oppositely charged pairs, where quantum mechanical
symmetrization effects do not contribute to the correlation
function \cite{Albe97}. 
Fig.~\ref{fig:PlusMinus}, panel (a), shows the correlation 
function for pairs
of oppositely charged
pions, obtained from the correlation of oppositely charged hadrons by 
applying a Monte-Carlo correction for non-pionic contaminations.
 They are plotted as a function of 

\begin{equation}
\label{eq:qinvdef}
Q_{inv} = \sqrt{(\Delta p_{x})^2 + (\Delta p_{y})^2 + 
(\Delta p_{z})^2 - (\Delta E)^2 }
\end{equation}
\noindent
near mid-rapidity and for a restricted $K_\bot$ range for central
collisions of S+Ag and Pb+Pb. Also shown is the
calculated correlation function for the point-like Gamow function
(solid line), and the correlation function due to Coulomb attraction
based upon finite source sizes of 4 fm (dotted line) and 6 fm (dashed
line), following the approach of Baym and Braun-Munzinger \cite{Baym96}. 
It is seen that the Coulomb attraction calculated for
a point-like source does not describe the measured data, and that the
finite source size needs to be taken into account.

Fig.~\ref{fig:PlusMinus}, panels (b) and (c), show the measured 
two-particle correlation function of negative 
pions, extracted from the negative hadrons with the same procedure mentioned
 above, as a function of
\qinv\ for Pb+Pb collisions, both uncorrected (open circles) and
corrected for Coulomb repulsion effects (filled circles). Panel (b) uses the Gamow factor 
for a point-like source whereas panel (c) employs a parametrisation of 
the opposite-charge correlation
function in panel (a).
The two correction procedures generate
different correlation functions, both in the peak width at low \qinv\
and in the asymptotic convergence to unity at high \qinv. We find that
the finite-size correction leads to a constant value of the correlation 
function at $Q_{inv}\ge60$ MeV/c (see inset of panel (c)), while
the ``standard'' Gamow correction results in a slope of the correlation 
function at large $Q_{inv}$ and leads to a systematic 
underestimate of  the source parameters for large dense pion 
sources\footnote{
Note that the Gamow function for pairs of like charged
particles in a very good approximation is equal to the inverse
Gamow function for pairs of unlike charged particles;
a slight difference is seen only for $Q_{inv}<10$~MeV/$c$. The same is true
also for the model of Baym and Braun-Munzinger~\cite{Baym96}, 
independent of the assumed particle source size.}.
All correlation functions discussed in the remainder of this paper
have been corrected for Coulomb repulsion using the finite-size
source correction as in Fig.~\ref{fig:PlusMinus}, panel (c), obtained
from the opposite-charge correlation function.

\section{Determination of Source Parameters from Correlation Functions}
\label{sec:sourceparams}

The data set used for the two particle
correlation analysis consists of 80,000 central Pb+Pb events. 
After tracking quality cuts and non-primary vertex track rejection
the average multiplicity for a central Pb+Pb collision within
$2.9<\ypi<5.5$ is about 200 negative hadrons. 
The resulting number of close pairs in this sample enables 
the analysis of correlation functions in the three relative
momentum projections $Q_0,\:Q_{\bot}$ and $Q_{\parallel}$, in bins in
\kperp\ of width 100 MeV/$c$ and bins in pair rapidity \ypipi\ of width
0.5, with small statistical error. 
An independent analysis of the data of VTPC2 and the MTPCs is presented here,
with the two data sets overlapping in the region $3.4<\ypi<4.4$, where the 
results agree within $\pm5\%$.
In the overlapping region the data was not averaged, but the low rapidity 
points at $\ypi < 3.9$  are from VTPC2 and the remainder are from the MTPCs.
Preliminary results of such a differential 
analysis of the correlation functions have been presented in \cite{Kadi96}.

\begin{figure}
\epsfig{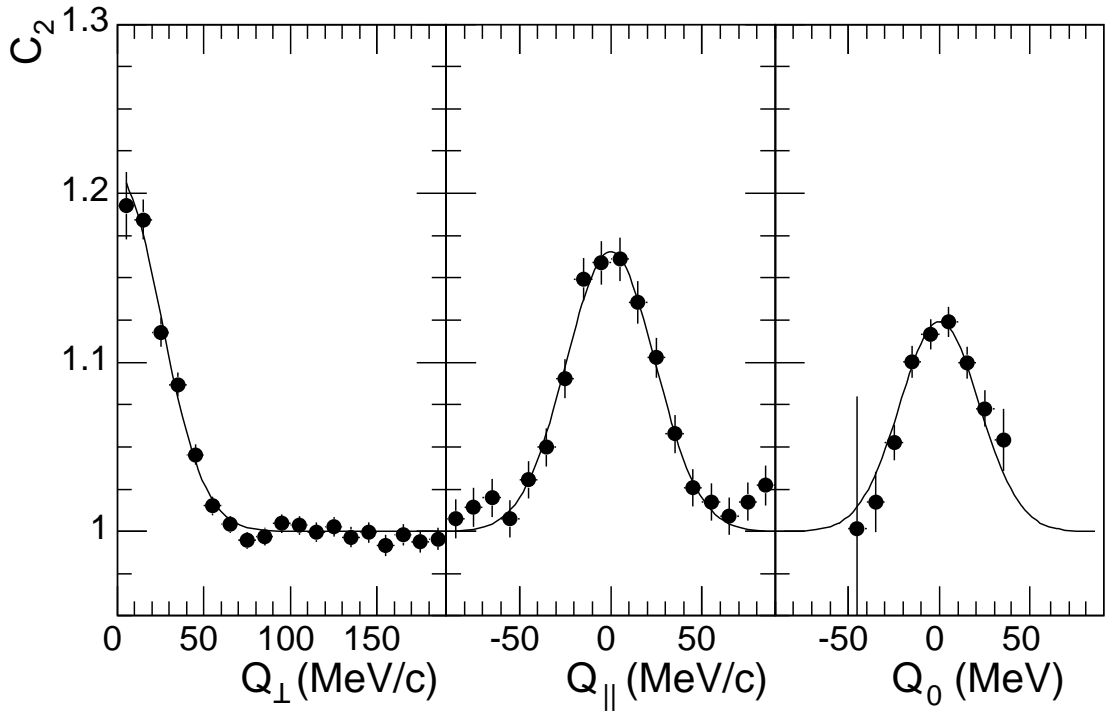}
\caption {
Two-particle correlation functions for negative hadrons from central
Pb+Pb collisions, as a function of $Q_{\bot}$, $Q_{\parallel}$ and
$Q_0$, in the interval $3.4<\ypipi<3.9$ and $0.1<\kperp<0.2$~GeV/$c$.
For each panel, the correlation function is integrated from 0
to 50 MeV/$c$ along the axes orthogonal to
the one plotted. Solid curves are the result of a 3-dimensional
maximum likelihood fit of equation (\ref{eq:ykph_corr_fn}) to the data
(see text).
\label{fig:CorrFn}
}
\end{figure}

The three dimensional fit to the expression
(\ref{eq:ykph_corr_fn}) 
is performed using the maximum likelihood
method to fit its five parameters,
with the correlation functions evaluated in the
Local Centre of Mass System which refers to the mean rapidity 
in each rapidity bin. 
Fig.~\ref{fig:CorrFn} shows a set of projections for a typical 
correlation function
of negative hadrons as a function of $Q_{\bot},\:Q_{\parallel}$ and
$Q_0$, within $3.4<\ypipi<3.9$ and $0.1<\kperp<0.2$~GeV/$c$. 
Also shown is the result of the fit with the 
parameters $R_{\bot} = 5.7$~fm, $R_{\parallel} = 7.4$~fm and
$R_0 = 3.5$~fm.
Non Gaussian source functions have been recently proposed, e.g.
\cite{Wied97}, but we find that a Gaussian source 
function fits the observed 
correlation functions at all rapidity and \kperp bins very well,
as shown in Fig.~\ref{fig:CorrFn}.

\begin{figure}
\begin{center}
\epsfig{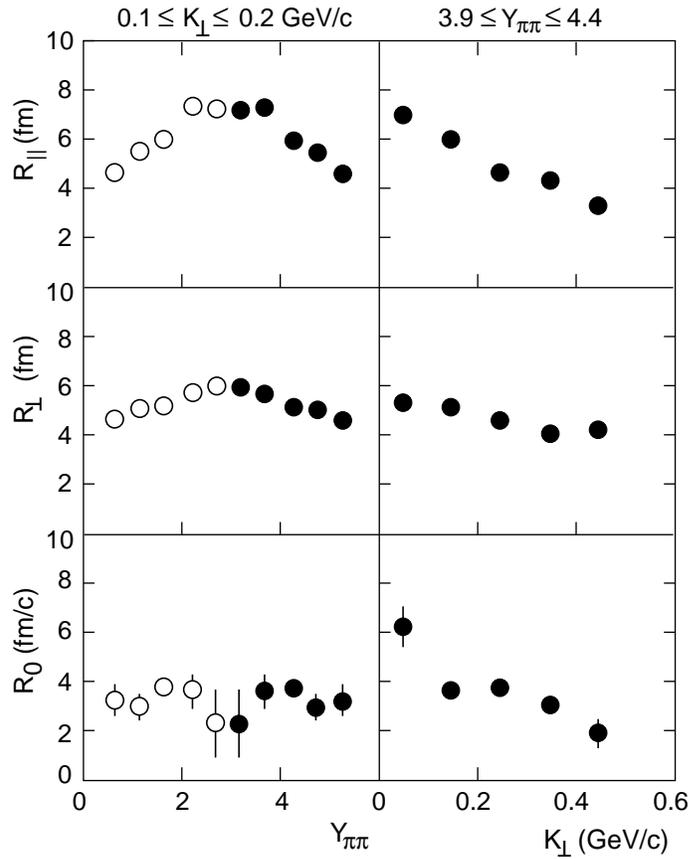}
\end{center}
\caption{
Examples of parameters \Rpar, \Rperp, and \Rzero\ resulting from a fit of
equation (\ref{eq:ykph_corr_fn}) to the two particle negative hadron
correlation functions for central Pb+Pb collisions.
Left panels:
parameters as a function of \ypipi\ within $0.1<\kperp<0.2$~GeV/$c$.
Filled circles are measured data, open circles are measured
data reflected about mid-rapidity (\ypipi= 2.9). Right panels:
parameters as a function of \kperp\ within $3.9<\ypipi<4.4$.
\label{fig:RVsK}
}
\end{figure}

Fig.~\ref{fig:RVsK} shows typical fit results for the parameters
$R_{\parallel},\:R_{\bot}$ and $R_0$ as a function of pair rapidity
\ypipi\ within $0.1<\kperp<0.2$~GeV/$c$ and as a function of \kperp\
within the interval $3.9<\ypipi< 4.4$.  At mid-rapidity
$(\ypipi= y_{cm} = 2.9)$, $R_{\parallel}$ and $R_{\bot}$ peak at
6-7~fm. The source appears to be locally isotropic. At higher
rapidities the source parameters are reduced to about 5.5~fm. The
parameter $R_0$, which reflects the duration of emission \cite{Hein96l,Hein96}, is
approximately constant at $R_0= \Delta \tau = 3-4$~fm/c. Previous
measurements of $\Delta\tau$ in two-particle correlation studies at
the SPS with Sulphur beams have yielded very small values of $R_0$, implying a
``sudden'' freezeout. 

We defer analysis of the correlation intensity parameter $\lambda$ to a
forthcoming publication but note here that throughout this analysis,
in which the pion pair purity is 55\%, the fit result for
$\lambda$ is about 0.45.

\begin{figure}
\begin{center}
\epsfig{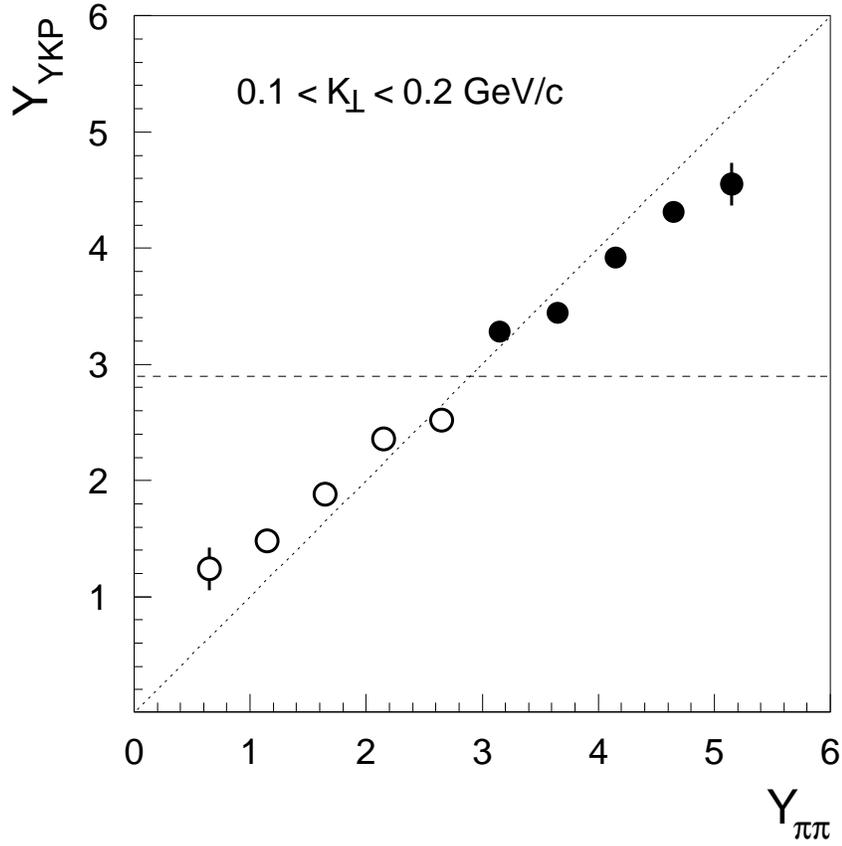}
\end{center}
\caption{
Yano-Koonin-Podgoretskii rapidity \yYKP\ vs. laboratory rapidity of the
pion pair \ypipi\ for central collisions of Pb+Pb. \yYKP\ is derived
from \Betayk\ (equation \ref{eq:yYKP}). Filled circles are data, open
circles are data reflected about mid-rapidity. The dashed horizontal line 
indicates
the correlation expected for a non-expanding source, the dotted line indicates
the correlation expected for an infinite boost-invariant source.
\label{fig:YykpVsYpi}
}
\end{figure}

\begin{figure} 
\begin{center}
\epsfig{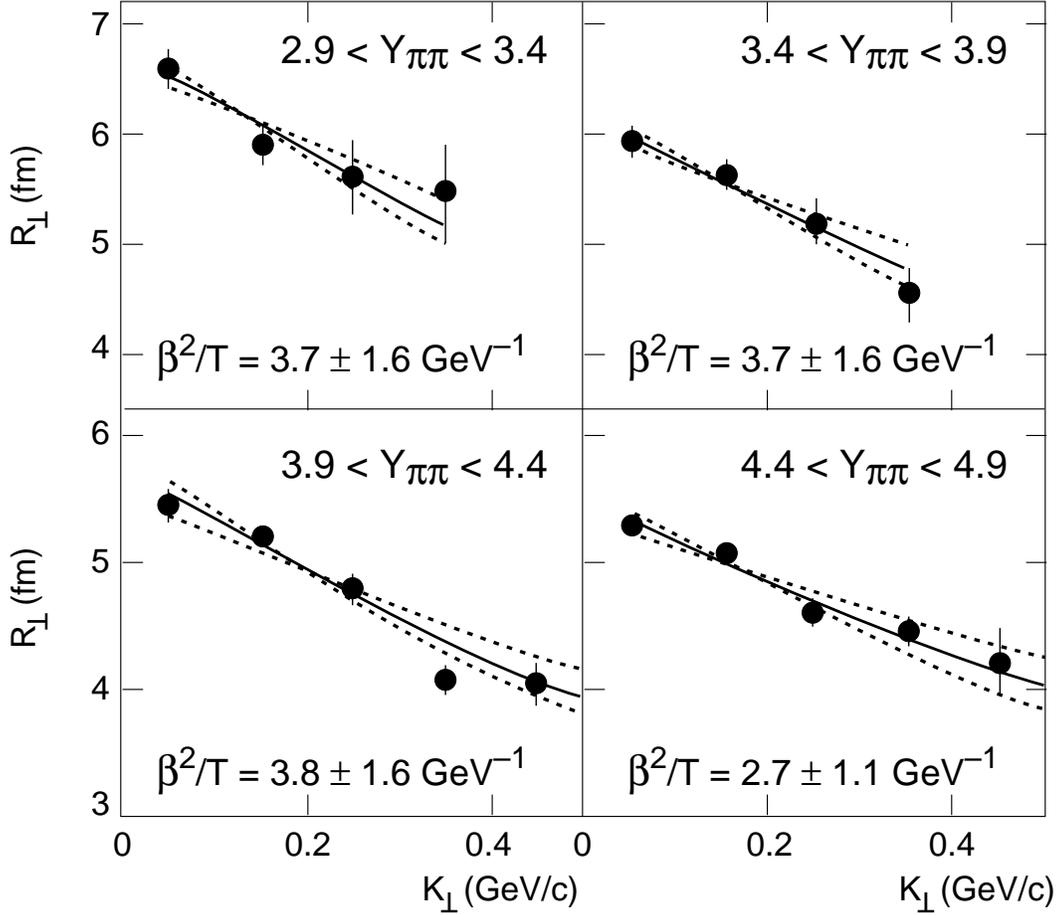} 
\end{center}
\caption{ 
Dependence of \Rperp\ on \kperp\ in four successive rapidity
intervals, for central Pb+Pb collisions. Solid curves show result
of the fit using equation (\ref{eq:heinzrperp}). Dashed curves show the
variation of the fit function when the $\beta_{\bot}^2/T$
parameter is varied one standard
deviation from the best fit values.
\label{fig:RVsKGlobal}
}
\end{figure}

Fig.~\ref{fig:YykpVsYpi} shows the dependence of the
``Yano-Koonin-Podgoretskii'' rapidity \yYKP\ as a function of pair
rapidity \ypipi\ for central Pb+Pb collisions. \yYKP\ is derived from
\Betayk\ as

\begin{equation}
\yYKP = {\frac{1}{2}}ln\frac{1+\Betayk}{1-\Betayk}+y_{cm}
\label{eq:yYKP}
\end{equation}

\noindent
where \Betayk\ is determined by fitting expression (\ref{eq:ykph_corr_fn}) 
to the negative hadron two-particle correlation functions
in five successive bins of pion pair rapidity \ypipi.
This representation was first used by the GIBS Collaboration \cite{Anik97}.
Fig.~\ref{fig:YykpVsYpi} shows that the rapidity \yYKP\ of the local
source is strongly correlated with the rapidity \ypipi\ of the emitted
pion pairs. A non-expanding source would exhibit no correlation,
whereas an infinite boost-invariant source would exhibit a strict
correlation.  The data are consistent with a boost-invariant source,
but show some deviation from the boost-invariant scenario at high
rapidity, as is expected for a beam of finite energy
($y_{beam}^{LAB} = 5.8$).
At \ypipi= 5 the longitudinal expansion velocity is found to be
\Betayk = 0.9.
We turn next to an analysis of the transverse source expansion, 
employing the dependence of \Rperp\ on \kperp\
which was predicted by Heinz \cite{Hein96}
to be sensitive to the presence and magnitude of a radial velocity field. 
Fig.~\ref{fig:RVsKGlobal} shows the dependence of \Rperp\ on \kperp\ 
for four different
bins of \ypipi, where \Rperp\ is determined by the fit of
(\ref{eq:ykph_corr_fn}) to the negative hadron two-particle
correlation functions. The solid line is the result of a fit to the
data by the function

\begin{eqnarray}
R_{\bot}\:=\:R_{G}\:[1\:+\:\frac{M_{T}\beta_{\bot}^2}{T}\:\cosh(\yYKP-\:\ypipi)]^{-1/2}\:;
\:\:\:\:\:\:M_T = \sqrt{m^2_{\pi}+K^2_{\bot}} 
\label{eq:heinzrperp}
\end{eqnarray}

This expression was derived by Heinz \cite{Hein96l} in a
model with longitudinal and transverse expansion, under the assumption
that the radial profile of the transverse velocity is given by
$\beta_{trans}(r)=\beta_{\bot}(r/R_{G})$,
where $R_{G}$ is the geometrical RMS radius of a Gaussian 
approximation to the
one dimensional transverse projection of the source density profile at 
freezeout, 
and $r$ is the distance from the centre
\footnote{In \cite{Wied97}
one notices that $\beta$ is replaced by a transverse rapidity $\eta$
in order to be valid in the relativistic regime. In the domain of $\beta$
we find in this study the differences between the two is of the order of 
our systematic error.}. 
The fit parameters are $R_{G}$ and the ratio $\beta_{\bot}^2/T$ with T the freezeout
 temperature. 
The dotted lines in
Fig.~\ref{fig:RVsKGlobal} show equation (\ref{eq:heinzrperp}) with
parameters one standard deviation away from those at minimum
$\chi^2$. The data in the three lowest rapidity bins are best
described with a common set of fit parameters: 
$R_{G}=(6.5\pm0.5)$~fm and
$\beta_{\bot}^2/T = (3.7 \pm1.6)$~GeV$^{-1}$. At forward rapidity both
parameters decrease, indicating that the pion emitting source is
ellipsoidal in shape.
We note here that the effect of long lived resonance decays on the \kperp 
dependence is ignored in the model leading to (\ref{eq:heinzrperp}). Wiedemann and 
Heinz \cite{Wied97} have shown, however, 
that this effect, in itself produces a slope of \Rperp  with \kperp  
which leads to a relatively small apparent value of
$\beta_{\bot}^2/T = 0.75$~GeV$^{-1}$ if (\ref{eq:heinzrperp}) is used to fit the 
\Rperp slope. 

\section{Discussion of Source Parameters}
\label{sec:sourceparamdisc}

The fit of equation (\ref{eq:heinzrperp}) to the \kperp dependence of 
\Rperp in Fig.~\ref{fig:RVsKGlobal} fixes the genuinely geometric transverse
source radius $R_{G}= 6.5$~fm near mid-rapidity.
In order to assess this result we recall the geometrical aspects 
of a Pb+Pb collision. We start from the $^{208}$Pb Woods-Saxon density 
distribution of radius
$R(^{208}$Pb) = 1.15$\times(208)^{1/3}$fm = 6.8~fm.
In an ideal zero impact parameter collision the primordial participant
baryons would be positioned in a cylindrical volume of this dimension.
However the NA49 trigger for central events selects a mean impact parameter 
of about 2~fm \cite{Albe95}, resulting in a slight reduction of the 
azimuthally averaged radius, to about $R_{participants} = 6$~fm. 
It is important, however, to consider next the transverse density distribution
in the primordial source.
Right after interpenetration, before significant transverse expansion has
occurred, this is a longitudinal cylinder.
The transverse density profile is obtained by projecting the two 
spherically uniform nucleon distributions onto a plane perpendicular 
to the beam ($z$) axis. 
In order to compare to the geometrical source parameter obtained
from the correlation analysis, we recall that the transverse radius 
$R_{G}$ = 6.5~fm
refers to a projection of the density profile of the cylinder onto one 
transverse dimension ($x$) only. 
Applying this projection also to the primordial
source cylinder, we obtain a nearly Gaussian density profile.
Overall, the two projections transform the shape of two uniform spheres
(where the radius is identical to the ``surface'') to approximately 
a Gaussian, with RMS radius smaller by $\sqrt{5}$ than the initial sphere 
radius. I.e. in the terminology employed for the source at freezeout, 
with RMS radius $R_{G}$ = 6.5~fm as obtained above, 
the primordial source has an RMS radius of 
$R_{prim} \approx 6/ \sqrt{5}$ fm = 2.7~fm.
The source thus expands by a factor of about 2.4 between initial maximum energy
density, and final hadronic decoupling.     
If it is assumed that the hadronic
transverse expansion velocity at the RMS radius position is 
$\beta \approx 0.5$  (see below), 
the system requires a  Gaussian mean expansion time of about
$\tau_{exp}$ = 8~fm/$c$.

The $K_{\bot}$ dependence of the source radius parameters shown in
Fig.~\ref{fig:RVsK} exhibits a steep decrease in $R_{\parallel}$ and a
shallower decrease of $R_{\bot}$.  The duration-of-emission parameter
$R_0$ stays roughly constant with perhaps some indication of a
decrease at high $K_{\bot}$. 
Heinz \cite{Hein96l} has provided an
estimate of $R_{\parallel}$ in the framework of the same source model that
leads to the estimate of $R_{\bot}$ employed above 
(equation (\ref{eq:heinzrperp})): 

\begin{eqnarray} R_{\parallel} =
{\tau}\:\left[{\frac {M_T}{T}}\:\cosh(\yYKP\:-\:\ypipi)\:-\: 
\frac {1}{\cosh^2(\yYKP-\ypipi)}\:+\:
\frac {1}{(\Delta\eta)^2}\right]^{-\frac{1}{2}} 
\label{eq:heinzrpar}
\end{eqnarray} 

where $\tau$ is the overall time of expansion, $T$
is the freezeout temperature which we take as 120~MeV and
$\Delta\eta$ is the width of the source in space-time rapidity 
which we take to be 1.3 (see bellow). 
We obtain $\tau \approx 8~$fm/$c$ near mid-rapidity, 
decreasing slightly to $\tau \approx 6$~fm/$c$
at high rapidity. Note that the overall expansion time $\tau$ of
the system can be longer than the interval $\Delta \tau$
during which hadrons decouple (freeze out) from the source (the
duration of the emission); indeed we observe that $\tau>\Delta
\tau= R_0$, where \Rzero\ is seen in Fig.~\ref{fig:RVsK} to be 3-4~fm/$c$.
Recalling the Gaussian parametrization of the correlation functions,
the temporal emission pattern of the hadronic fireball can thus be
pictured as a Gaussian with mean $\tau = 8$~fm/$c$ and $\sigma = R_0 =
\Delta \tau = 3.5$~fm/$c$.
This is consistent with our above estimate of
$\tau_{exp}$, the net duration of transverse expansion.
The emission of pions fades away at about 3$\sigma$, i.e. after 
8 + 3$\times$3.5 fm/$c$ $\approx $18~fm/$c$.

\section{Transverse Expansion and Single Particle Spectra}
\label{sec:spectra}

\begin{figure}
\epsfig{figure=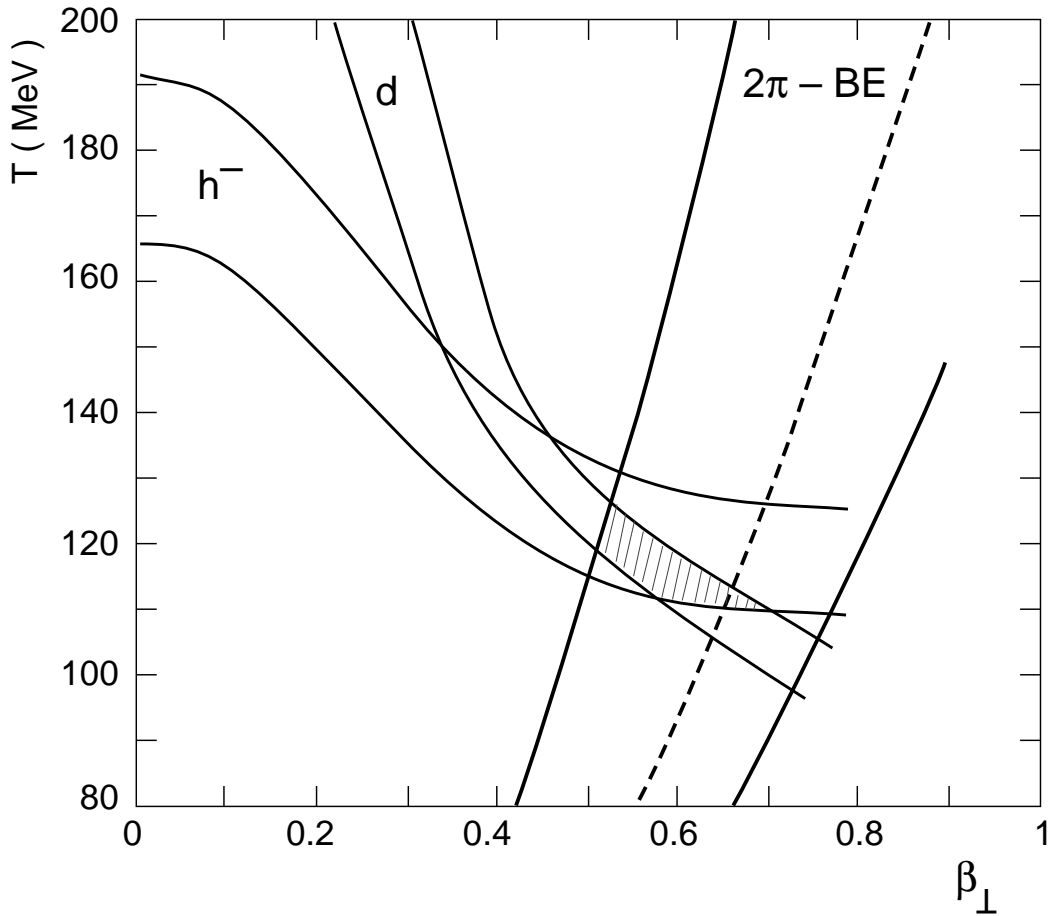,width=14cm}
\caption{
Allowed regions of freezeout temperature vs. radial velocity 
for central Pb+Pb collisions near mid-rapidity, derived from the
fit of equation (\ref{eq:heinzrperp}) to negative hadron two-particle
correlation functions and from negative hadron and deuteron transverse mass
spectra fitted by expression (\ref{eq:dndpt}). Bands are drawn at $\pm\sigma$ 
around fitted values.
\label{fig:TVsBeta}
}
\end{figure}

The $K_{\bot}$ dependence of $R_{\bot}$ extracted from the two-particle 
correlation data and the fits
using equation (\ref{eq:heinzrperp}) were shown in Fig.~\ref{fig:RVsKGlobal} 
for four rapidity bins. 
The first three rapidity bins are best described with a common set of
parameters, $R_{G}= (6.5\pm0.5)$~fm and 
$\beta_{\bot}^2/T = (3.7 \pm1.6)$~GeV$^{-1}$.
Because the fit parameter $\beta_{\bot}^2/T$ is a ratio, 
there is an ambiguity in the transverse velocity $\beta_{\bot}$ and
freezeout temperature $T$, ranging from $\beta_{\bot}$ = 0.86 at
$T$ = 200~MeV to  $\beta_{\bot}$ = 0.61 at $T$ = 100~MeV 
(see Fig.~\ref{fig:TVsBeta}).
The implied average transverse energy of a single pion varies dramatically over
this domain. We remove this ($T,\beta_{\bot}$)
ambiguity 
and enforce transverse energy conservation by 
fitting the NA49 transverse mass
spectra \cite{Afanas} within the same model. 

It has been
shown \cite{Bear96,Dodd95,PBM95,NA44b} that the spectra of pions and heavier
hadrons in Pb+Pb collisions are consistent with the assumption of a
radial velocity field. 
We employ here a parametrization given by Chapman et al.
\cite{Chap95} to fit the single particle transverse mass distribution for
central Pb+Pb collisions. 
It refers to the space-time source function \cite{Hein96l} as was employed in our above
analysis of pion correlations
(equation~(\ref{eq:heinzrperp}).
The single particle distributions in this case can be approximated by the 
expression

\begin{eqnarray}
\label{eq:dndpt}
P_1(m_T,y) \propto m_T R_{*}^2 \Delta \eta_{*}
\left[1+ \frac{R_{*}^2}{2R_{G}^2}(\Delta \eta)_{*}^2- 
\frac{m_T}{8T}(\Delta \eta)_{*}^4 \right]* \\ \nonumber
* \exp{ \left[ -\frac{m_T}{T} + \frac{\beta_{\bot}^2 (m_T^2
- m^2)}{2T(T+m_T \beta_{\bot}^2)} - \frac{ 0.5 y^2}{ ( \Delta \eta)^2 + T/m_T}
\right] }
\end{eqnarray}
where
\begin{equation}
\frac{1}{R_{*}^2} = \frac{1}{R_{G}^2}\left(1+\beta_{\bot}^2\frac{m_T}{T} \right)
\mbox{\quad and \quad}
\frac{1}{(\Delta \eta)_{*}^2} = \frac{1}{(\Delta \eta)^2} + \frac{m_T}{T}
\mbox{\quad with \quad}
m_T = \sqrt{m^2_{\pi}+ p^2_{T}}
\end{equation}

As already mentioned above in the context of equation (\ref{eq:heinzrperp}), 
the model's parameters are the temperature $T$, 
the transverse geometrical source radius $R_{G}$, 
the transverse expansion velocity $\beta_{\bot}$ at the distance $R$
and the width $\Delta \eta$ of the source in the space-time rapidity.
The latter is related to the width of the rapidity distribution  $\Delta y$
by
 
\begin{equation}
(\Delta y)^2 = (\Delta \eta)^2 + \frac{T}{m_T}
\end{equation}

\noindent For negative hadrons in a realistic temperature range $T$ from 80 
to 180 MeV,
$\Delta y = 1.4$  and $<m_{T}> = 0.45$~GeV giving $\Delta \eta = 1.3 \pm 0.1$. 

The fit
results in again an ambiguous set of solutions with regard to $T$ and
$\beta_{\bot}$, but now with a different constraint than the two pion
correlation, namely the observed mean transverse momentum.
For negative hadrons at mid-rapidity we thus obtain ``true''
temperatures at freezeout ranging from $T\approx180$~MeV at
$\beta_{\bot} = 0$ to $T\approx120$~MeV at $\beta_{\bot}\approx0.7$. 

The allowed regions
in the plane of freezeout temperature $T$ versus transverse
surface velocity $\beta_{\bot}$ that result both from the correlation and 
spectral analysis of negative hadrons are shown in Fig.~\ref{fig:TVsBeta}. 
Overlayed on the same figure 
is the analogous region of allowed $T$ versus $\beta_{\bot}$ 
obtained by NA49 from the measured transverse mass spectrum of 
deuterons \cite{NA49sp}
which were identified by dE/dx
and time-of-flight. The contribution of the background under the deuteron peak
was determined in small bins of total and transverse momentum \cite{Str97} and 
has been subtracted.
The bands shown correspond to 
a one $\sigma$ variation from the best fit values
taking into account the systematic errors. For negative hadrons 
the choice of the fit interval gives a contribution to the systematic 
error which accounts for half of the 
width of the band, for the deuterons this is negligible.
A similar analysis of (more preliminary) NA49 data was carried out
by K\"ampfer \cite{Kamp96}.

All data employed in Fig.~\ref{fig:TVsBeta} refer to 
a rapidity domain about 0.6 units away from
mid-rapidity. The negative hadron spectra and correlation data refer to
$3.4<y<3.9$, the deuteron spectra to $2<y<2.4$ (kinematically
equivalent relative to \ycm= 2.9). The details of the analysis of
hadronic spectra will be given in a forthcoming publication \cite{NA49sp}.

From Fig.~\ref{fig:TVsBeta} we see that the three independent results 
favour a narrowly defined overlap region, $T= (120 \pm 12)$~MeV and
$\beta_{\bot}= 0.55\pm0.12$.
The errors given here contain an estimate of the systematic error of the negative hadron 
and deuteron spectral data included in the width of the bands in Fig.~\ref{fig:TVsBeta}.
The remaining systematic error from the correlation data was estimated to be
negligible for T and to be about (+0.02, -0.05) for $\beta_{\bot}$. 
No correction has been made for the decay of
long-lived resonances. However the deuteron data should be fairly
free of such effects, and Heinz has shown \cite{Hein96} that the
$K_{\bot}$ dependence of $R_{\bot}$ is only weakly affected by this at
the relatively high radial velocity extracted here.

\section{Conclusions}
\label{sec:conclusions}

We have carried out a high statistics analysis of Bose Einstein
correlations of negative pions in central
Pb+Pb collisions at 158 GeV per nucleon, 
which has allowed a differential study of the dependence
of the correlation function on the pair kinematic variables, 
in small bins of rapidity and transverse momentum of the pair.
The correlation between opposite-charge hadrons measured 
simultaneously in the same 
apparatus is used to experimentally determine the Coulomb 
correction, a method 
that is shown to be superior to the correction with the standard Gamow factor.
 
The representation of the correlation function derived by Heinz \cite{Hein96l,Hein96} which is most suitable 
for a source with longitudinal collective motion is employed
to extract the parameters for an expanding source. 
The five fitted parameters  $\lambda,\:R_{\bot},\:R_{\parallel},\:R_0$ and 
$\beta_{YK}$ are functions of the pair rapidity \ypipi and
transverse momentum \kperp. The duration of emission given by 
$R_0$  is 3 to 4 fm/c showing that there is a finite
duration of the hadronic freezeout. 
It is nevertheless small and not consistent with 
a long duration of a mixed (hadronic-partonic) phase implied by a strong 
first order phase transition \cite{Bert89}. 
This statement might have to be modified in a scenario with significant source
expansion before hadronization as was pointed out already by Pratt \cite{Bert89}.
The rapidity dependence of $\beta_{YK}$ is found to correspond to a source 
with large longitudinal expansion, close to the boost-invariant limit.

We then employ a model which contains longitudinal and transverse 
expansion \cite{Hein96l}
to extract information about the transverse dynamics of the source 
from the \ypipi and \kperp dependence of the transverse radius $R_{\bot}$.
The results correspond to a
source with a Gaussian radius $R_{G}= (6.5 \pm0.5)$~fm, 
which is a factor of 2.4
larger than the initial transverse RMS radius of the high density zone. 
The transverse dynamics are indicated by the parameter 
$\beta_{\bot}^2/T= (3.7 \pm1.6)$~GeV$^{-1}$, which however 
cannot fix the transverse velocity and the temperature of the system
independently.
The use of the transverse momentum spectra of deuterons and negative hadrons
give further constraints fixing  $T= (120 \pm 12)$~MeV and
$\beta_{\bot}= 0.55\pm0.12$.

With the temperature of the source fixed, the rapidity dependence of 
$R_{\parallel}$
gives the overall expansion time $\tau= 8$ fm/c. This agrees with the 
expansion 
time needed for the system to grow transversely by a factor of 2.4 with a 
transverse velocity $\beta_{\bot}$= 0.55, showing the consistency of the 
results
of the longitudinal and transverse expansion modes.

The observation of longitudinal expansion is not unique to central 
nuclear collisions, and 
follows from the longitudinal dynamics of incompletely stopped
partons and the decay of their ``strings'' predominantly ordered in beam 
direction \cite{Beca97}.
However, no trivial explanation is at hand for the origin(s) of the large
transverse collective velocity field. It may arise from the combined 
effects of the initial hadronic and partonic scattering cascades \cite{Geig97} 
(in the extreme case from the transverse expansion of a 
QCD partonic state prior to hadronization),
or from an isentropic expansion of the tightly packed hadrons
following hadronization, that either generates a radially ordered velocity 
field, or enhances such an initial field if it exists already at 
hadronization \cite{Fere96,Herm95}.\\

\vspace{0.5cm}
\noindent
Acknowledgements: This work was supported by the Director, 
Office of Energy Research, 
Division of Nuclear Physics of the Office of High Energy and Nuclear Physics 
of the US Department of Energy under Contract DE-ACO3-76SFOOO98, 
the US National Science Foundation, 
the Bundesministerium fur Bildung und Forschung, Germany, 
the Alexander von Humboldt Foundation, 
the UK Engineering and Physical Sciences Research Council, 
the Polish State Committee for Scientific Research (2 P03B 01912), 
the EC Marie Curie Foundation,
the Polish-German Foundation
and the Hungarian Scientific Research Foundation under the 
contracts T1492 and T7330.


\begin{thebibliography}{99}
\bibitem{Laer96} E. Laermann, Nucl. Phys. A610 (1996) 1c, 
                 and references therein.
\bibitem{Albe95} T. Alber et al., NA49 Collaboration,
                 Phys. Rev. Lett. 75 (1995) 3814.
\bibitem{Gold60} G. Goldhaber, S. Goldhaber, W. Lee and A. Pais, Phys. 
                 Rev. 120 (1960) 300; G. Cocconi, Phys. Lett. 49B (1974) 459.
\bibitem{Shur80} E.V.Shuryak, Phys. Rep. 61 No.2 (1980) 72.
\bibitem{Bjor83} J. D. Bjorken, Phys. Rev. D27 (1983) 140.

\bibitem{Siny89} A. N. Makhlin and Y. M. Sinyukov, Z. Phys. C39 (1988) 69;
                 Y. M. Sinyukov, Nucl. Phys. A498 (1989) 151.
\bibitem{Bert89} G. F. Bertsch, Nucl. Phys. A498 (1989) 173;
                 S. Pratt, T. Cs\"org\"o and J. Zimanyi, 
                 Phys. Rev. C42 (1990) 2646;
                 S. Pratt, Phys. Rev. D33 (1986) 1314.

\bibitem{Hein96l} U. Heinz et al., Phys. Lett. B382 (1996) 181;
                 S. Chapman, J. R. Nix and U. Heinz, Phys. Rev. C52 (1995) 2694.
\bibitem{NA35a}  T. Alber et al., NA35 Collaboration,
                 Phys. Rev. Lett. 74 (1995) 1303;
                 T. Alber et al., NA35 Collaboration, Z. Phys. C66 (1995) 77.
\bibitem{NA44a}  A. Franz et al., NA44 Collaboration, 
                 Nucl. Phys. A610 (1996) 240c, and references therein.
\bibitem{Bach91} J. B\"achler et al., NA35 Collaboration, 
                 Z. Phys. C52 (1991) 239.
\bibitem{Schn93} E. Schnedermann, J. Sollfrank and U. Heinz, 
                 Phys. Rev. C48 (1993) 2462.

\bibitem{Fere96} D. Ferenc, Nucl. Phys. A610 (1996) 523c.

\bibitem{Herm95} M. Herrmann and G.F. Bertsch, Phys. Rev. C51 (1995) 328.

\bibitem{Bear96} I. G. Bearden et al., NA44 Collaboration,
                  Nucl. Phys. A610 (1996) 175c.

\bibitem{Afanas} S. V. Afanasiev et al., NA49 Collaboration, 
                 Nucl. Phys. A610 (1996) 188c.

\bibitem{Hage83} R. Hagedorn Riv. Nuov. Cimento 6 No.10 (1983) 1.

\bibitem{Dodd95} J. Dodd et al., NA44 Collaboration, 
                 Nucl. Phys. A590 (1995) 523.

\bibitem{Soll94} J. Sollfrank, M. Gazdzicki, U. Heinz and J. Rafelski, Z. Phys.
                 C61 (1994) 659, and references therein.

\bibitem{Albe97} T. Alber et al., NA35 Collaboration, Z. Phys. C73 (1997) 443.

\bibitem{YKP78}  F. B. Yano and S. E. Koonin, Phys. Lett. B78 (1978) 556;
                 M. I. Podgoretskii, Sov. J. Nucl. Phys. 37 (1983) 272.

\bibitem{Hein96} U. Heinz, Nucl. Phys. A610 (1996) 264c.

\bibitem{Appe97} H. Appelsh\"auser, PhD thesis Frankfurt 1997;
                 S. Sch\"onfelder, PhD thesis, MPI Munich 1997 (MPI-PhE/97-09).


\bibitem{Wern88} K. Werner, Phys. Rev. D 39 (1989) 780; 
                 Z. Phys. C 42 (1989) 85.

\bibitem{Brun15} R. Brun, et al., GEANT 3.15 CERN user's manual.


\bibitem{Baym96} G. Baym and P. Braun-Munzinger, Nucl. Phys. A610 (1996) 286c.

\bibitem{Kadi96} K. Kadija, NA49 Collaboration, Nucl. Phys. A610 (1996) 248c.

\bibitem{Wied97} U. Wiedemann and U. Heinz, Phys. Rev. C56 (1997) R610.

\bibitem{Anik97} M. Kh. Anikina et al., GIBS Collaboration, 
                 Phys. Lett. B397 (1997) 30.

\bibitem{PBM95}  P. Braun-Munzinger, J. Stachel, J. P. Wessels and N. Xu, 
                 Phys. Lett. B344 (1995) 43; Phys. Lett. B365 (1996) 1.

\bibitem{NA44b}  I. G. Bearden et al., NA44 Collaboration, 
                 Phys. Rev. Lett. 78 (1997) 2080.

\bibitem{Chap95} S. Chapman, P. Scotto and U. Heinz, Heavy Ion Physics
                 1 (1995) 1.

\bibitem{NA49sp} NA49 Collaboration, to be published.


\bibitem{Str97}  C. Struck, Diplom thesis Marburg 1997. 


\bibitem{Kamp96} B. K\"ampfer, Rossendorf preprint FZR-149 (1996).

\bibitem{Beca97} F. Becattini, M. Ga\'zdzicki and J. Sollfrank,  
                 hep-ph/9710529, to be published in Z. Phys. C.

\bibitem{Geig97} K. Geiger and D. K. Srivastava, preprint nucl-th/9706002.

\end{thebibliography}
\end{document}